\documentclass[%
preprint,
 amsmath,amssymb,
 aps,
]{revtex4-2}

\usepackage{graphicx}% Include figure files
\usepackage{dcolumn}% Align table columns on decimal point
\usepackage{bm}% bold math
\usepackage[dvipsnames]{xcolor}
\newcommand{\ts}[1]{\textcolor{red}{#1}}

\begin{document}

\preprint{APS/123-QED}

\title{Enhanced robustness of evolving systems with bipartite topology}
%Enhanced Robustness of Evolving Open Systems with Bipartite Network
% Force line breaks with \\

\author{Fumiko Ogushi}%
 \email{fumiko.ogushi@mi.meijigakuin.ac.jp}
\affiliation{Faculty of Mathematical Informatics, Meiji Gakuin University\\
1518 Kamikurata-cho, Totsuka-ku, Yokohama-shi, Kanagawa, 244-8539 JAPAN}

\author{Kimmo Kaski}
\affiliation{Aalto University School of Science, P.O. Box 11000 (Otakaari 1), FI-00076 AALTO, Finland}

\author{Takashi Shimada}
\affiliation{
Mathematics and Informatics Center, The University of Tokyo\\
\&
Department of Systems Innovation, Graduate School of Engineering, The University of Tokyo,
7-3-1 Hongo, Bunkyo-ku, Tokyo, 113-8656 Japan
}
\email{shimada@sys.t.u-tokyo.ac.jp}
  

\date{\today}

\begin{abstract}
%\ts{
%We investigate the robustness of \kk{open} systems with bipartite network topology under successive introduction of new entities. Analogous to conventional unstructured systems, bipartite systems exhibit a phase transition between diverging and finite phases. When the initial degrees of newly introduced elements in the two partitions are asymmetric, the transition point shifts to larger degree, indicating enhanced robustness against additions of new elements. This effect becomes stronger with increasing asymmetry, and strikingly, the diverging phase persists even when both initial degrees exceed the critical point of the corresponding unstructured system. A mean-field analysis reveals the physical origin of this robustness: in the asymmetric regime, the mean degree of the emergent network can be significantly larger than the initial degree, which suppresses extinctions while extinct elements remain biased toward recently introduced low-degree nodes. These combined effects uncover a simple and universal mechanism \kk{that governs the} robustness of evolving asymmetric bipartite systems.}
% (2026/3/31: 148 words < PRX Limit 150 words (limit for PRL & PRE is ~300 words)
Evolving open systems, in which new entities are continually introduced and those turning unfit go extinct, exhibit a phase transition between a diverging phase, where the system size grows indefinitely, and a finite phase, where it remains bounded. We show that imposing a bipartite interaction topology alone leaves this transition unchanged when the two partitions are introduced with equal initial connectivity. In contrast, when the initial degrees are asymmetric, the robustness of the system is markedly enhanced such that the transition shifts to higher connectivity and the diverging phase persists even when both initial degrees individually exceed the critical point of the corresponding unstructured system. In addition, we find a re-entrant transition, i.e. a return to the diverging phase as asymmetry is increased while the initial degree of one of the partitions %at one side 
is fixed, making it lying entirely outside the original mean-field picture. An extended mean-field analysis identifies the 
origin of these effects such that in the asymmetric regime, a feedback between the bipartite handshaking constraint and
different extinction rates drives the mean degree of emergent network far above the initially assigned connectivity. This degree elevation suppresses extinction probabilities across the community while simultaneously concentrating extinctions among recently introduced, low-degree nodes. The interplay of these two effects constitutes a simple and universal robustness mechanism for evolving systems with asymmetric bipartite structure.
\end{abstract}

%Use showkeys class option if keyword
\maketitle

%\tableofcontents

\section{Introduction}
Many natural and man-made systems with two types of interacting entities can be considered to have %as having 
a bipartite structure~\cite{GuillaumeLatapy2006, Neal2024}, 
like the %e.g., 
plant-pollinator systems~\cite{Bascompte2007, DeokSunLee_PRL2012, TMKnight_Science2013, Wei_Nature2021}, microbial communities %in a 
interacting through resources and metabolites~\cite{DSouzaKost_NatProdRep2018, YamagishiSaitoKaneko2021, MoranTikhonov_PhysRevX2022, Clegg-Gross_PNAS2025, Shimada2026arXiv},
social networks~\cite{Quayle2025, Cremers2025},
the Wikipedia community of editors and articles~\cite{Oogushi_WikipediaEcology2021SREP, Shimada_WikipediaModel2023PhysicaA},
and the country-product network in world trades~\cite{Hidalgo_2009PNAS, Tacchella_2012SREP, Tacchella_2018NaturePhysics, JT_CDcriteria2026arXiv}.
Most of these systems are going through an evolutionary process of having new species (entities) appearing by invasion or mutation and loss of species by extinction.
The stability of these systems against such large disturbances with the change in degree of freedom is of primary interest for understanding the origin and 
mechanism of sustaining the complexity of the system.

A recent bipartite network based study %\kk{bipartite studies} %formulations 
of microbial communities has highlighted how the discrete dependency structure constrains their diversity~\cite{Clegg-Gross_PNAS2025}, but its explanatory power %their explanatory power 
is largely confined to the resulting equilibrium. By contrast, models of population dynamics with ordinary differential equations let us track temporal trajectories, yet are often computationally costly, especially for studying large disturbances. A key missing framework is therefore one that retains discrete low-dimensional node states while explicitly capturing the propagation of changes in time. Discrete network dynamical models provide this bridge, preserving simplicity while enabling transient dynamics, cascade timing, and path-dependent approaches to community reorganization. In fact, a simple graph-dynamics-based framework has previously been shown to capture essential features of the stability of systems with random interaction~\cite{Takashi_EOS_SREP2014} and systems with bidirectional interactions~\cite{Fumiko_EOS_SREP2017}.

%Here, we will focus on investigating the robustness of systems with a bipartite network topology in their interaction, under successive introduction of new entities based on the dynamical graph approach. As in many real systems the two parts are rarely symmetric, we will investigate systems in which the two parts have different initial degrees. We find that the systems with such asymmetry become more robust against the the inclusion of a species with random interaction. This results in a wider parameter regime for the system to grow in size. The effect of asymmetry in the interactions will be described on the basis of previous studies on systems without bipartite structure.
%KK: THIS FOLLOWING PARAGRAPH IS SUGGESTED TO SUBSTITUTE THE PARAGRAPH ABOVE:
%TS on 26th June: OK!

Here we extend the dynamical graph framework to systems with strict bipartite interaction topology and focus on investigating their robustness under successive introduction of new species when the two partitions carry initial asymmetric connectivity. We show that such an asymmetry generically enhances robustness, shifting the phase boundary well beyond the predictions of a symmetric mean-field theory. We observe a re-entrant transition to the diverging phase for large asymmetry, which is a phenomenon that is absent in the unstructured case. We also demonstrate that both effects arise from a degree-shift that is intrinsic to the bipartite structure, as asymmetric input connectivity drives the emergent network to a self-consistent state in which mean degrees deviate substantially from their initial values, and extinctions become concentrated among newly introduced nodes. An extended mean-field theory incorporating these effects quantitatively reproduces the observed phase boundaries, including the re-entrant transition.

This paper is organized such that next we briefly describe the model in Section II followed in Section III showing results for the expected behavior of the bipartite model, phase diagram for symmetric and asymmetric cases, and shift in the degrees of the emergent networks. In Section IV we present concluding remarks.

%%% We have confirmed that the sentences till here is complete! (TS & KK, 2026/7/1)

\section{Model}
We introduce a model in which species are grouped into two types, $\alpha$ and $\beta$. The interaction among species has a strict bipartite topology, i.e., species do not have interaction with other species in the same group but interact with some of the species in the other group. The interactions are represented by a directed link from species $j$ to species $i$ with a weight of the link being $a_{ij}$. We calculate the fitness of each species $i$ by the simple sum of the weights of its incoming links:
\begin{equation}
	f_i = \sum_{j} a_{ij}.
\end{equation}
The species with positive fitness ($f_{i}>0$) can survive and therefore remain in the community.
If the fitness of a species is non-positive ($f_{i}\leq 0$), that species goes extinct and the node it represents is removed with the links to and from that species. That may in turn cause the extinction of other species. Such cascading-extinctions are checked by re-calculating the fitness of other species. If we have two or more species with non-positive fitness, we first remove the least fit species and re-evaluate the fitness of the other species.

If the fitness of the species are all positive, the state of the system is persistent (or stable).
Every time after getting to such persistent state, we add a new species to the system. The type of the new species $\chi$ is randomly chosen from the group $\alpha$ or $\beta$ with equal probability $1/2$. Then we assign $m_\chi$ (i.e. $m_\alpha$ or $m_\beta$) to new interactions (links with random directions and random weights). The direction of each link (from or to the species) is chosen randomly with probability $1/2$ and the random weights are drawn from the normal distribution $N(0, 1)$. After the inclusion of the new species, we repeat the re-calculation of the fitness and check for extinctions until a new persistent state is reached.
We repeat this ``evolutionary-assembly'' process
%\kk{for the system to get to the state of robustness}
to access the robustness of the system against the successive introduction of new species. 

\begin{figure}[htbp]
\includegraphics[width=1.0\linewidth]{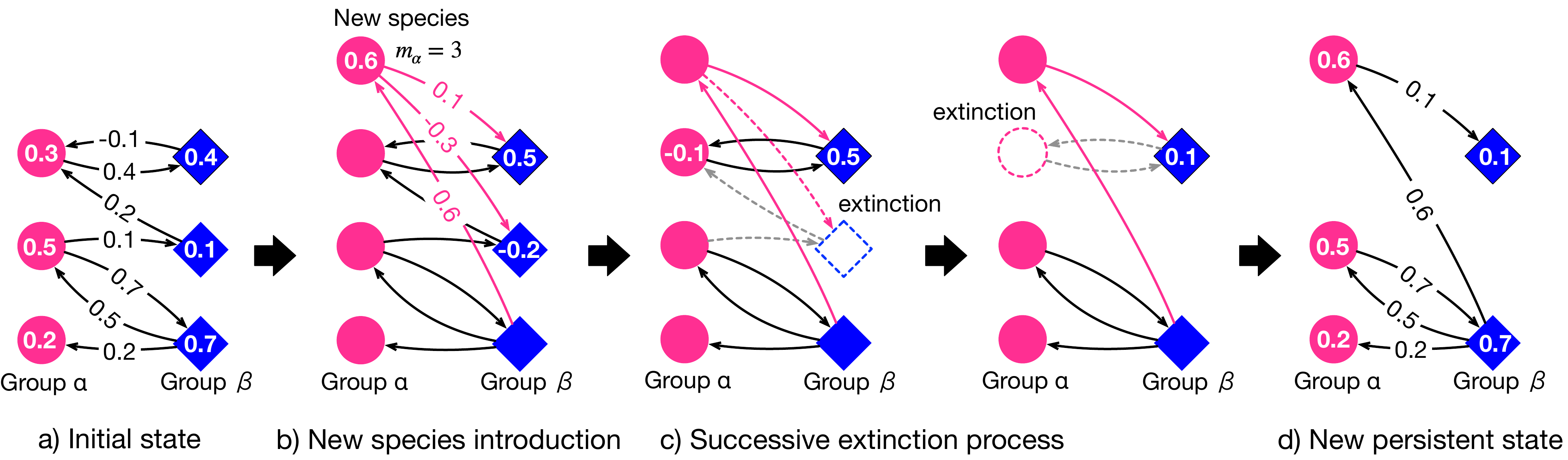}
\caption{The %evolving 
bipartite graph model of two groups and its discrete temporal evolution.
(a) Initial persistent state, with all species having positive fitness.
(b) Introduction of a new species. The type of new species is randomly selected from either Group $\alpha$ or Group $\beta$. Then, according to the assigned initial degree $m_\alpha$ ($m_\beta$), $m_\alpha$ ($m_\beta$) random links from the new species to randomly chosen species in the other group are formed. %are formed to the new species. 
(c) Successive extinction process. The inclusion of a new species
can cause extinction(s) of
%causes an extinction of a
resident species and that extinction may trigger %triggers
another extinctions of a resident species.
(d) Next persistent state the system has reached.}%. Finally, the system reaches a new persistent state.}
\label{fig_model}
\end{figure}

\section{Results}
\subsection{Expected behavior of the bipartite model}
Let us estimate the behavior of the current model based on the original non-structured model. In the original model, the system shows the transition from the finite phase (the number of nodes fluctuates within a finite range) to the diverging phase (system size slowly diverges with time) as as the unique input parameter $m$ increases.
The basic mechanism of this transition can be captured by a mean-field analysis. In this theory, supported by the observation in systematic simulation, we treat the emerging system as a random network with infinitely large size and average degree $k = m$~\cite{Takashi_EOS_SREP2014}.
Under this approximation, the average number of extinctions $N_E$ per newly introduced species is calculated as a simple infinite geometric series of the average probability of extinction of the species in the community against the single link addition or deletion event $E$, as
\begin{equation}
N_E = \sum_{l=1}^{\infty}\left(\frac{mE}{2}\right)^{l} = \frac{mE}{2-mE},
\end{equation}
where $mE/2$ is the direct impact of the inclusion or extinction of species on neighboring resident species.
The critical point for growth in the number of species corresponds to $N_E = 1$. 
Therefore, we obtain a simple condition for the transition point, 
\begin{equation}
 m_c E(m_c) = 1.
\end{equation}
Because $mE$ is an increasing function of $m$ while $E(m)$ is a decreasing function of $m$, the mean-field treatment correctly predicts the phase transition from the diverging phase to the finite phase~\cite{Takashi_EOS_SREP2014}.

Let us apply this straightforward mean field approach %theory 
to the current setting, by %with 
assuming that the average degree on each side is also kept near the input degree, i.e. $k_\alpha = m_\alpha, k_\beta = m_\beta$.
For the current bipartite network model, we denote the direct impact of an inclusion or extinction of a species at each side (the average number of extinctions on the nearest neighbors) as
\begin{equation}
    I_\alpha(m_\alpha, m_\beta) = \frac{m_\alpha E(m_\beta)}{2}, \
    I_\beta(m_\alpha, m_\beta) = \frac{m_\beta E(m_\alpha)}{2},
\end{equation}
respectively, where $E(m)$ is the average probability that a resident species goes extinct at a link addition or deletion event.
We also denote the mean and geometric mean of those impacts as
\begin{equation}
    \mu_I(m_\alpha, m_\beta) = \frac{I_\alpha(m_\alpha, m_\beta) + I_\beta(m_\alpha, m_\beta)}{2}, \ \gamma_I(m_\alpha, m_\beta) = \sqrt{I_\alpha(m_\alpha, m_\beta) I_\beta(m_\alpha, m_\beta)},
\end{equation}
respectively.
Then the expectation value of the net number of extinctions on each side triggered by the inclusion of species of type $\alpha$ and $\beta$ with probability $p_\alpha$ and $p_\beta$ are calculated as
\begin{equation}
    \begin{pmatrix}
        N_E^{\alpha}\\
        N_E^{\beta}
    \end{pmatrix}
    =
    \begin{pmatrix}
        N_E^{\alpha\alpha} & N_E^{\alpha\beta}\\
        N_E^{\beta\alpha} & N_E^{\beta\beta}
    \end{pmatrix}
    \begin{pmatrix}
        p_\alpha\\
        p_\beta
    \end{pmatrix},
\end{equation}
where the elements of the matrix $N_E^{\chi\eta}$ denote the expectation value of the number of extinctions in group $\chi$ triggered by the inclusion of a new species in $\eta$, 
are calculated as
\begin{eqnarray}
N_E^{\alpha \alpha}(m_\alpha, m_\beta)
&=&
I_\alpha I_\beta + I_\alpha^2 I_\beta^2 + I_\alpha^3 I_\beta^3 + \cdots
=
\frac{\gamma_I^2}{1 - \gamma_I^2},
\\
N_E^{\alpha \beta}(m_\alpha, m_\beta)
&=&
I_\beta + I_\alpha I_\beta^2 + I_\alpha^2 I_\beta^3 + \cdots
=
\frac{I_\beta}{1 - \gamma_I^2},
\\
N_E^{\beta \alpha}(m_\alpha, m_\beta)
&=&
I_\alpha + I_\alpha^2 I_\beta + I_\alpha^3 I_\beta^2 + \cdots
=
\frac{I_\alpha}{1 - \gamma_I^2},
\\
N_E^{\beta \beta}(m_\alpha, m_\beta)
&=&
I_\alpha I_\beta + I_\alpha^2 I_\beta^2 + I_\alpha^3 I_\beta^3 + \cdots
=
\frac{\gamma_I^2}{1 - \gamma_I^2}.
\end{eqnarray}
For the current equal inclusion probability $p_\alpha = p_\beta = 1/2$, the average number of extinctions after each inclusion event reads as follows
\begin{equation}
N_E(m_\alpha, m_\beta)
=
\frac{N_E^{\alpha\alpha} + N_E^{\alpha\beta} + N_E^{\beta\alpha} + N_E^{\beta\beta}}{2}
=
\frac{\mu_I + \gamma_I^2}{1 - \gamma_I^2},
%=
%\frac{\mu_I(m_\alpha, m_\beta) + \gamma_I^2(m_\alpha, m_\beta)}{1 - \gamma_I^2(m_\alpha, m_\beta)},
\end{equation}
and the condition for the phase boundary is found again at $N_E = 1$.

For systems with symmetric initial degree, $m_\alpha = m_\beta = m$, this condition yields
the same equation as the original system~\cite{Takashi_EOS_SREP2014}:
\begin{equation}
N_E(m, m)
= \frac{\mu_I}{1 - \mu_I}
= \frac{m E(m)}{2 - mE(m)}.
\end{equation}
Therefore, as long as $mE$ is an increasing function of $m$, the mean-field treatment predicts the same phase transition from diverging phase to finite phase at the same point as the original system $m_\alpha = m_\beta = m_c$.

\begin{figure}[tbhp]
\includegraphics[width=0.7\linewidth]{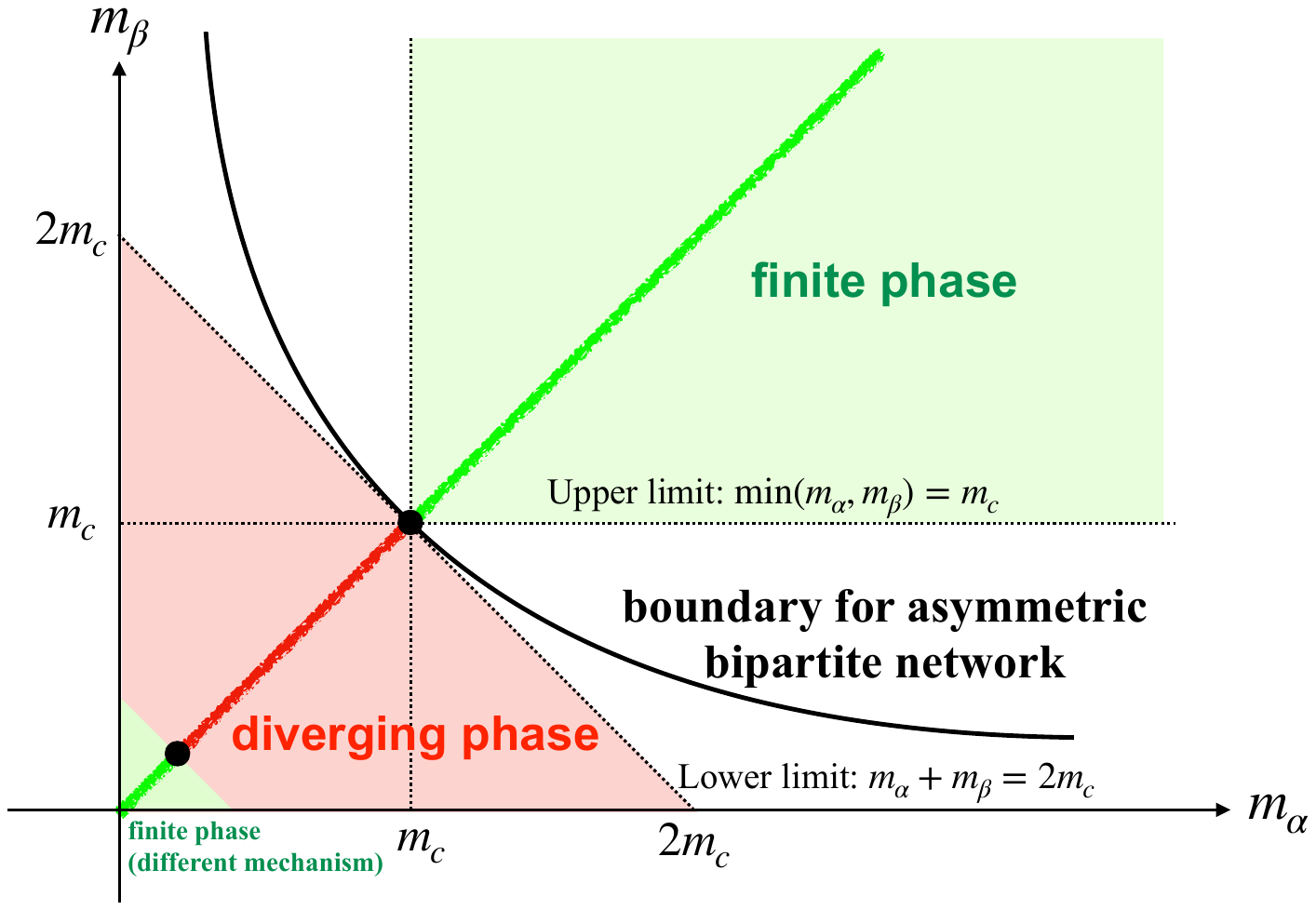}
\caption{
%Schematic
The expected phase diagram of %portrait of the 
a system with bipartite topology. The phase boundary in the initial degree ($m_\alpha, m_\beta$) - plane predicted by %from 
the %\ts{straightforward} 
mean field theory should run within the white (non-shaded) region and has %with 
concave shape, as shown by the solid curve. The phase boundary is bounded above and below by the critical point $m_c$ of the symmetric system ($m_\alpha = m_\beta$).
}
\label{fig_MF_PhasePortrait}
\end{figure}

For the asymmetric case, $m_\alpha \neq m_\beta$,
%with losing no generality,
the estimated average number of extinctions becomes
\begin{equation}
N_E(m_\alpha, m_\beta)
=
\frac{\mu_I}{1 - \mu_I} \left( \frac{1 - \mu_I}{1 - \gamma_I} \cdot \frac{1 + \frac{\gamma_I^2}{\mu_I}}{1 + \gamma_I} \right).
\end{equation}
Using the general inequality $\mu_I \ge \gamma_I$ and the monotonicity of $E$ and $mE$: $\frac{{\rm d}}{{\rm d} m} E < 0$ and $\frac{{\rm d}}{{\rm d}m} (mE) > 0$,
we can find the relation
\begin{equation}
N_E(m_\alpha, m_\beta) < \frac{\mu_I}{1 - \mu_I} = N_E(\mu_m, \mu_m)
\quad \left( \mu_m = \frac{m_\alpha + m_\beta}{2} \right).
% We can add the left inequarity but I omitted it for simplicity (TS)
%\mu_I < N_E^{m_\alpha > m_\beta} < \frac{\mu_I}{1 - \mu_I} = N_E^{m_\alpha = m_\beta}.
\end{equation}
This means that the asymmetric bipartite system can be %is
more robust compared to the symmetric system or non-bipartite system with the same average degree.
We can also find an upper limit for this reinforcement effect of asymmetry,
$\min(m_\alpha, m_\beta) \le m_c$,
from the inequality $\frac{{\rm d}}{{\rm d} m_\alpha} N_E(m_\alpha, m_\beta) > 0$ which is again derived from monotonicity $\frac{{\rm d}}{{\rm d} m} E < 0 \cap \frac{{\rm d}}{{\rm d}m} (mE) > 0$. Therefore, in total, we obtain a schematic structure of the expected phase boundary of the bipartite system between the diverging phase and the finite phase, as shown in FIG~\ref{fig_MF_PhasePortrait}.

% comment out for simplicity
\if{0}
A better representation of the shape of the phase boundary can be made by taking the linear expansion of $I_\chi$ around the critical point as $I_\chi (m_c + \chi) = \frac{1}{2} + c_1 \chi +\cdots$, which yields the phase boundary line $N_E(m_\alpha, m_\beta) = 1$ as
\begin{equation}
    m_\beta = m_c - \frac{m_\alpha - m_c}{1 + \frac{c_1}{3} (m_\alpha - m_c)}
    \qquad
    (m_\alpha > m_\beta).
    \label{Eq_PhaseBoundary_LinearExpansion}
\end{equation}
\fi
% comment out for simplicity

\subsection{Phase diagram}
\subsubsection{Symmetric case ($m_\alpha = m_\beta = m$)}
\begin{figure}[tbhp]
\includegraphics[width=0.7\linewidth]{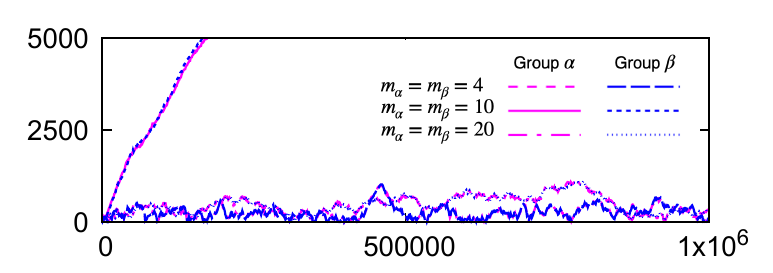}
\caption{Temporal evolutions of the total number of species in Group $\alpha$ and Group $\beta$. The system size diverges if the initial degree of each new species $m_\alpha = m_\beta = m$ is moderately sparse ($m=10$), and otherwise keeps fluctuating in a finite range ($m=4$ and $m=20$).}
\label{fig_n_symm}
\end{figure}

\begin{figure}[tbhp]
\includegraphics[width=0.7\linewidth]{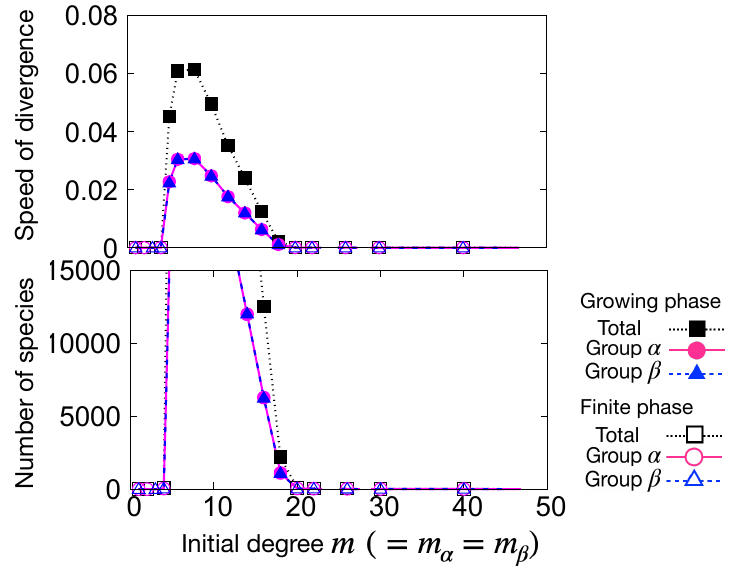}
\caption{The growth behavior of emergent system with symmetric initial degree, $m_\alpha = m_\beta = m$.
Filled symbols correspond to the diverging phase, in which the speed of divergence for Group $\alpha$ and Group $\beta$, $\lim_{t \to \infty} N_{\alpha}(t)/t$ and $\lim_{t \to \infty} N_{\beta}(t)/t$, are equal or larger than $0.0006$, which is to empirically ensure thei positiveness.
%i.e. $\lim_{t \to \infty} > 0.0006 \ \cap \ \lim_{t \to \infty} \frac{N_{\beta}(t)}{t} \geq 0.0006$.
This symmetric bipartite system show a transition from the growing phase to the finite phase at around $m=18$, which is consistent with the mean-field prediction $m_c = m_c^{\rm{original}} = 18.5$.
}
\label{fig_v_symm}
\end{figure}

We first investigate the case of symmetric initial degree, i.e., $m_\alpha = m_\beta = m$. In this case, we can confirm the prediction from the mean-field argument that having a bipartite topology makes no difference, as shown in Fig.~\ref{fig_n_symm} and Fig.~\ref{fig_v_symm}.
%%%%% Symmetric case: Temporal dynamics
%%%%% Symmetric case: Number of species & velocity
%%%

%\newpage 
\subsubsection{Asymmetric case ($m_\alpha \neq m_\beta$)}
\begin{figure}[tbhp]
\includegraphics[width=0.6\linewidth]{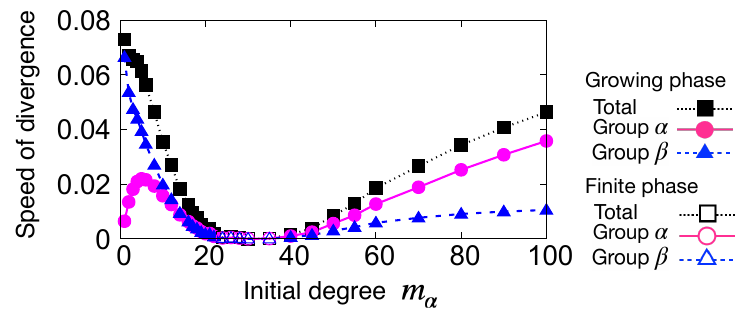}
\caption{ 
%\ts{
Speed of divergence, 
$\lim_{t \to \infty} N_\alpha(t)/t$ and 
$\lim_{t \to \infty} N_\beta(t)/t$, of emergent systems for the initial degree of Group $\alpha$ is $1 \leq m_{\alpha} \leq 100$, while keeping the initial degree of Group $\beta$ fixed at $m_{\beta}=16$. The temporal behaviors of the number of species in the both groups coherently show a transition from growing phase to finite phase at $m_{\alpha} \sim 24$, and then show a re-entrant transition back to the growing phase at $m_{\alpha} \sim 40$. %} 
}
\label{fig_v_mb16}
\end{figure}

\begin{figure}[htbp]
\includegraphics[width=0.6\linewidth]{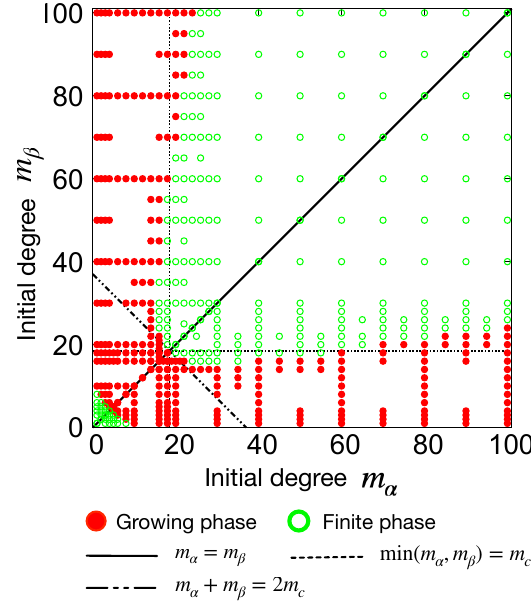}
\caption{Phase diagram of the bipartite model for asymmetric initial degrees. The data for $m_\alpha < m_\beta$ are mirrored from $m_\alpha > m_\beta$.
Filled circles represent the growing phase ($\lim_{t \to \infty} N_\alpha / t \geq 0.0006 \ \cap \ \lim_{t \to \infty} N_\beta(t) / t \geq 0.0006$) and open circles represent the finite phase ($\lim_{t \to \infty} N_\alpha(t) / t < 0.0006 \ \cup \ \lim_{t \to \infty} N_\beta(t) / t < 0.0006$).
Dotted and double-dotted lines represent the upper and lower limit for the boundary estimated from the mean-field theory, respectively. %The solid curve shows the phase boundary estimated by the mean-filed approximation with the linear expansion of $I$ around the critical point (Eq. (\ref{Eq_PhaseBoundary_LinearExpansion})), the linear coefficient of which is obtained from the original model.
}
\label{fig_pd}
\end{figure}

Next, we investigate the case in which the initial degrees of the newly introduced species are asymmetric, i.e., $m_\alpha \neq m_\beta$. This produces a phase diagram that is qualitatively different from the mean-field prediction (FIG.~\ref{fig_pd}) 
as well as demonstrated in FIG.~\ref{fig_v_mb16} by the section in $m_\beta = 16$. For this sweep, we see that the system sizes for both types coherently show a transition from the growing phase to the finite phase around $m_{\alpha} = 20$, which is consistent with the mean-field prediction. However, as the system increases further $m_\alpha$, it %as further increasing $m_\alpha$, the system 
shows a re-entrant transition back to the growing phase around $m_{\alpha} = 45$, which is not possible in the mean-field picture. 

%%%%% Asymmetric case: Speed of divergence for 10<m1<100, fixing m2=16
%%

%%%%% Phase diagram

\clearpage

\subsection{A drastic shift in the degrees of emergent networks and its impact}
We have found that the actual phase boundary is surprisingly different from what is expected from a naive extension of formally valid mean field theory. In the following we will show that this difference stems from the strong structuring of the emergent systems, namely, the average degree.

As shown in FIG.~\ref{fig_k_mb16_w_estimation}, the average degree of an emergent network can considerably deviate from the initial degrees $m_\alpha$ and $m_\beta$, when those are not symmetric. This is in contrast to the original model with no bipartite topology, in which the degree of the emergent network is always close to the initially given degree.
Let us estimate the cause and %the 
effects of this difference.

\subsubsection{The cause and the condition for the degree shift of emergent networks}
For emergent networks with $N_\alpha$ and $N_\beta$ nodes in the $\alpha$ and $\beta$ groups, respectively, the relation $N_\alpha k_\alpha = N_\beta k_\beta$ must hold for the average degrees of each side $k_\alpha$ and $k_\beta$ due to %from a 
trivial ``handshaking condition''.
In the divergent phase, in which both groups increase linearly in time with growth rates $g_\alpha$ and $g_\beta$, %therefore, 
the actual degrees of the emergent network should obey
\begin{equation}
\frac{k_\alpha}{k_\beta} = \frac{N_\beta}{N_\alpha} \sim \frac{g_\beta t}{g_\alpha t} = \frac{g_\beta}{g_\alpha}.
\label{eq_handshake}
\end{equation}
This means that the ratio between the degrees of the two groups is not directly related to $m_\alpha$ and $m_\beta$.
%, in the vicinity of the phase boundary, i.e., $g_\chi \ll 1$.

\begin{figure}[tbhp]
\includegraphics[width=0.7\linewidth]{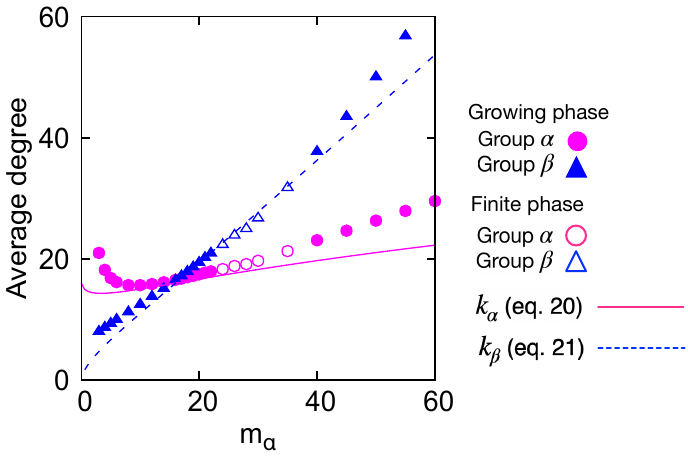}
\caption{
Average degrees of emergent system for changing the initial degree $m_{\alpha}$, while  
keeping the initial degree of Group $\beta$ %$m_{\beta}$ 
fixed at $m_\beta = 16$. 
Symbols represent the average degrees of emergent system in Group $\alpha$, $k_{\alpha}$, (magenta circles) and in Group $\beta$, $k_{\beta}$, (blue triangles). 
Solid and dashed lines represent the theoretical predictions for $k_\alpha$ and $k_\beta$, which are calculated from Eqs.~(\ref{eq_ka_estimation}) and (\ref{eq_kb_estimation}) using $\nu = 2/3$.
}
\label{fig_k_mb16_w_estimation}
\end{figure}

In the finite phase and in the growing phase near the phase boundary, i.e. $g_\chi \ll 1.$ 
The condition that direct (and hence the major) impact of inclusions of new species in each group to the other, 
\begin{equation}
    \tilde{I}_\alpha = \frac{m_\alpha E(k_\beta)}{2},
    \
    \tilde{I}_\beta = \frac{m_\beta E(k_\alpha)}{2},
\end{equation}
should valance, with an approximate empirical functional form of
\begin{equation}
    E(k) = c k^{-\nu}
    \quad
    \left( c \approx \frac{1}{2}, \ \nu \approx \frac{2}{3} \right)
\end{equation}
leads to the relation
\begin{equation}
    \left( \frac{k_\beta}{k_\alpha} \right)^\nu
    =
    \frac{m_\alpha}{m_\beta}.
\end{equation}
Coupling this with another simple relation required for the balance between the average gain and loss of the links of the system (i.e. average degrees are constant in time),
\begin{equation}
m_\alpha + m_\beta = k_\alpha + k_\beta,
\label{eq_gain_loss}
\end{equation}
we obtain
\begin{eqnarray}
  k_\alpha
  &=&
  \frac{m_\alpha + m_\beta}{1 + \left(m_\alpha/m_\beta \right)^\nu}
  = \left(\frac{m_\alpha + m_\beta}{m_\alpha^\nu + m_\beta^\nu}\right)m_\beta^\nu,
%  = \left(\frac{m_\alpha + m_\beta}{2}\right) \left[ \frac{2}{1 + \left(m_\alpha/m_\beta \right)^\nu} \right],
\label{eq_ka_estimation}
  \\
  k_\beta
  &=&
  \frac{m_\alpha + m_\beta}{1 + \left( m_\alpha/m_\beta \right)^{-\nu}}
   = \left(\frac{m_\alpha + m_\beta}{m_\alpha^\nu + m_\beta^\nu}\right)m_\alpha^\nu.
%  = \left( \frac{m_\alpha + m_\beta}{2} \right) \left[ \frac{2}{1 + \left(m_\alpha/m_\beta \right)^{-\lambda}} \right].
    \label{eq_kb_estimation}
\end{eqnarray}
Note that this relation corresponds to an ``equilibrium'', since $k_\alpha$ and $k_\beta$ deviated from this should feel a negative feedback, via the shift in the population ratio $N_\alpha/N_\beta$ and the ``hand-shaking relation''.
This estimation explains why the average degrees of emergent network deviate from the initially given degrees when those are not symmetric, and why $k_\beta$ grows faster than $k_\alpha$ as we increase $m_\alpha$ while keeping %with fixing 
$m_\beta$ fixed (FIG.~\ref{fig_k_mb16_w_estimation}).

%%%%% Estimated degree by Equation 26 -> To Be 19 (2026/3/11)

\subsubsection{The impact of the degree shift to the system}
As we have seen above, the actual mean degrees of the emerging system $k_\alpha$ and $k_\beta$ in the current bipartite model can be considerably shifted from that of the newly introduced species, $m_\alpha$ $m_\beta$. Let us examine the impact of this shift on the robustness of the emerging system.

Under the degree shift, the direct impacts of an inclusion or extinction of a species on each side are as follows
\begin{eqnarray}
    \tilde{I}_\alpha(m_\alpha, m_\beta; k_\alpha, k_\beta) &=& \frac{m_\alpha E(k_\beta)}{2} = \frac{m_\alpha}{k_\alpha} \cdot I_\alpha(k_\alpha, k_\beta),
    \\
    \tilde{I}_\beta(m_\alpha, m_\beta; k_\alpha, k_\beta) &=& \frac{m_\beta E(k_\alpha)}{2} = \frac{m_\beta}{k_\beta} \cdot I_\beta(k_\alpha, k_\beta),
\end{eqnarray}
respectively, where the extinction probability function $E(x)$ is assumed to be the same as the original one. Since the further impact is calculated from the propagation on the emergent network, the average net number of extinctions triggered by an inclusion of species of type $\alpha$ and $\beta$ are calculated as
\begin{eqnarray}
\tilde{N}_E^{\alpha \alpha}(m_\alpha, m_\beta; k_\alpha, k_\beta)
&=& \tilde{I}_\alpha I_\beta + \tilde{I}_\alpha I_\alpha I_\beta^2 + \tilde{I}_\alpha I_\alpha^2 I_\beta^3 + \cdots
= \left( \frac{m_\alpha}{k_\alpha} \right) N_{\alpha\alpha}(k_\alpha, k_\beta),
\label{eq_tildeNaa}
\\
\tilde{N}_E^{\alpha \beta}(m_\alpha, m_\beta; k_\alpha, k_\beta)
&=& \tilde{I}_\beta + \tilde{I}_\beta I_\alpha I_\beta + \tilde{I}_\beta I_\alpha^2 I_\beta^2 + \cdots
= \left( \frac{m_\beta}{k_\beta}  \right) N_{\alpha\beta}(k_\alpha, k_\beta),
\\
\tilde{N}_E^{\beta \alpha}(m_\alpha, m_\beta; k_\alpha, k_\beta)
&=& \tilde{I}_\alpha + \tilde{I}_\alpha I_\alpha I_\beta + \tilde{I}_\alpha I_\alpha^2 I_\beta^2 + \cdots
= \left( \frac{m_\alpha}{k_\alpha} \right) N_{\beta\alpha}(k_\alpha, k_\beta),
\\
\tilde{N}_E^{\beta \beta}(m_\alpha, m_\beta; k_\alpha, k_\beta)
&=& \tilde{I}_\beta I_\alpha + \tilde{I}_\beta I_\beta I_\alpha^2 + \tilde{I}_\beta I_\beta^2 I_\alpha^3 \cdots
= \left( \frac{m_\beta}{k_\beta} \right) N_{\beta\beta}(k_\alpha, k_\beta).
\label{eq_tildeNbb}
\end{eqnarray}
The dash-dotted lines in FIG.~\ref{fig_improved_mortalities} shows these extended mean field theory estimates %theoretical estimations,
to that of emergent systems (filled circles). We find that the actual number of extinctions in the %on 
$\alpha$ group per inclusion of new species into it, %on $\alpha$ group, 
$n_E^{\alpha \alpha}$, is considerably smaller than the mean-field estimation $\tilde{N}_{\alpha\alpha}$, while other three cases show good agreement between the theory and the observation. This difference is the remaining problem which hinders %prevents
us to explain the re-entrance to growing phase at around $m_\alpha = 40$ by the mean-field condition
\begin{equation}
\tilde{N}_E (m_\alpha, m_\beta; k_\alpha, k_\beta)
=
\frac{1}{2} \left( \tilde{N}_E^{\alpha\alpha} + \tilde{N}_E^{\alpha\beta} + \tilde{N}_E^{\beta\alpha} + \tilde{N}_E^{\beta\beta} \right) = 1.
\end{equation}

\clearpage
\subsubsection{Strong reinforcement effect due to the structuring in group $\beta$.}
\begin{figure}[tbhp]
\includegraphics[width=1.0\linewidth]{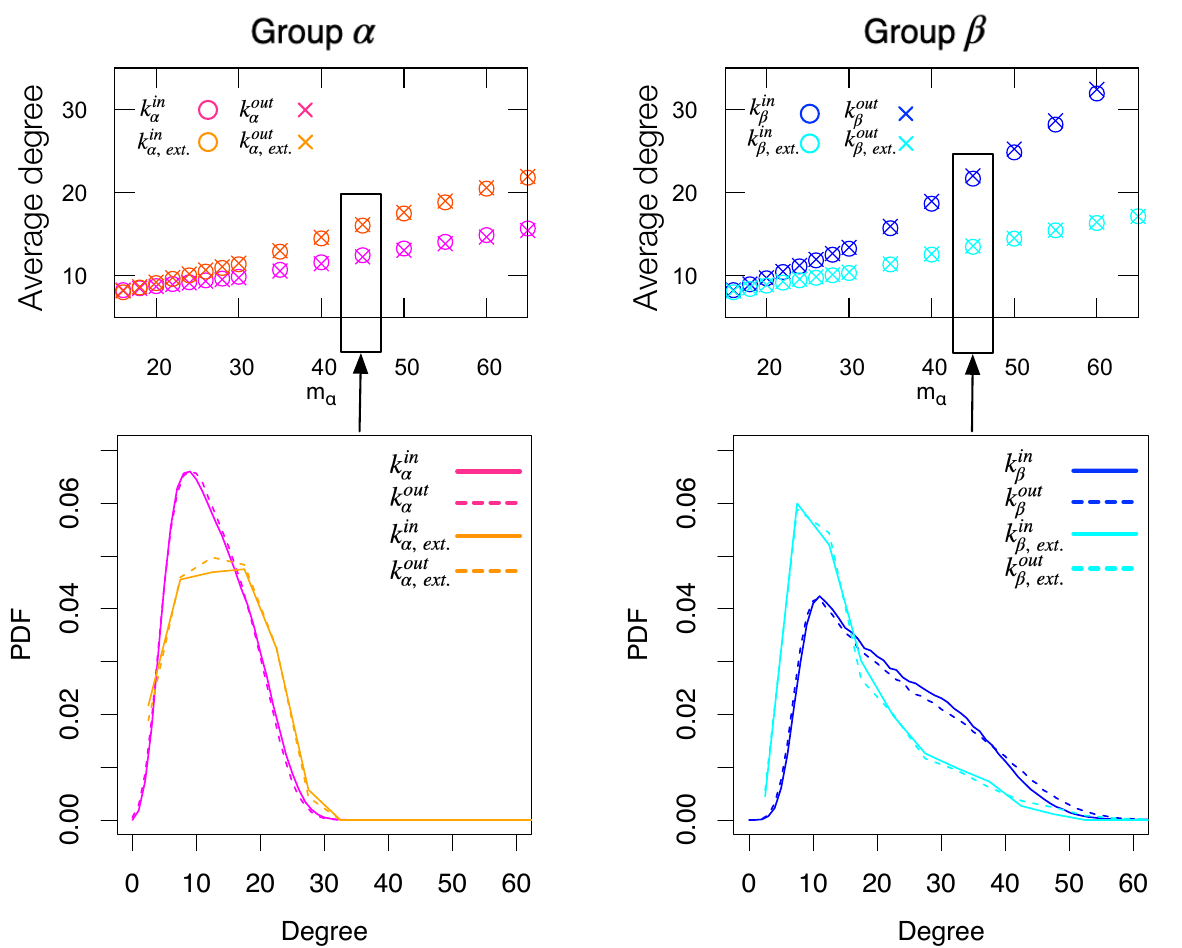}
\caption{
%\ts{
(Top): The change in the average degrees in each group as a function of $m_\alpha$, while keeping $m_\beta$ fixed at $16$.
The deviation of the degrees of species going to extinct from the community average is larger in Group $\beta$.
(Bottom): The distributions of species in the emergent network and that of species going to extinction, obtained from the system in the ``re-entrant'' growing phase (with the initial degrees $m_\alpha = 45$ and $m_\beta = 16$). The in- and out- degrees are drawn by solid and dashed lines, respectively.
%}
}
\label{fig_kextinct}
\end{figure}

The overestimation of $n_E^{\alpha\alpha}$ suggests a further reinforcement mechanism on $\beta$ group. Indeed, in this group, we find that the average degree of the species going to extinct is considerably smaller than the average of entire population: $k_\beta^{ext.} \approx m_\beta < k_\beta$ (FIG.~\ref{fig_kextinct}).
This happens basically because the average degree of the emergent network is larger than the initially given degree.
This %That 
means that the species which has experienced more link change events (extinction or introduction of species connecting to that node) tends to get larger fitness and hence becomes more robust. So, the species
that are going extinct in Group $\beta$ are 
%\kk{is likely to get} %goes extinct in %such a group %are more 
dominated by the new-comers that have smaller amount of in- and out-going links.
This difference is negligible in the %other 
Group $\alpha$, where the average degree of the emergent network is not significantly greater than the degree initially given. The difference is also negligible in the systems with a smaller asymmetry in the initial degrees.

%%%%% Degrees of extinct species

Knowing the difference, we should replace the estimate %estimation
by treating the extinction probability functions for {\it species introduction} $E_i$ and {\it species deletion} $E_e$ on both sides separately.
\begin{eqnarray}
\check{N}_E^{\alpha \alpha}
&=&
\left( \frac{m_\alpha E_i^\beta}{k_\alpha E_e^\beta} \right)
\left( \check{I}_\alpha \check{I}_\beta + \check{I}_\alpha^2 \check{I}_\beta^2 + \check{I}_\alpha^3 \check{I}_\beta^3 + \cdots \right)
=
\left( \frac{m_\alpha E_i^\beta}{k_\alpha E_e^\beta} \right)
\left( \frac{\gamma_{\check{I}}^2}{1-\gamma_{\check{I}}^2} \right),
\label{eq_checkNaa}
%=
%\left( \frac{m_\alpha E_i(k_\beta)}{k_\alpha E_e(k_\beta)} \right) N_{\alpha \alpha},
\\
\check{N}_E^{\alpha \beta}
&=&
\left( \frac{m_\beta E_i^\alpha}{k_\beta E_e^\alpha} \right)
\left( \check{I}_\beta + \check{I}_\alpha \check{I}_\beta^2 + \check{I}_\alpha^2 \check{I}_\beta^3 + \cdots \right)
=
\left( \frac{m_\beta E_i^\alpha}{k_\beta E_e^\alpha} \right)
\left(\frac{\check{I}_\beta}{1-\gamma_{\check{I}}^2} \right),
\label{eq_checkNab}
%=
%\left( \frac{m_\beta E_i(k_\alpha)}{k_\beta E_e(k_\alpha)} \right) N_{\alpha \beta},
\\
\check{N}_E^{\beta \alpha}
&=&
\left( \frac{m_\alpha E_i^\beta}{k_\alpha E_e^\beta} \right)
\left( \check{I}_\alpha + \check{I}_\alpha^2 \check{I}_\beta + \check{I}_\alpha^3 \check{I}_\beta^2 + \cdots \right)
=
\left( \frac{m_\alpha E_i^\beta}{k_\alpha E_e^\beta} \right)
\left( \frac{\check{I}_\alpha}{1-\gamma_{\check{I}}^2} \right),
\label{eq_checkNba}
%=
%\left( \frac{m_\alpha E_i(k_\beta)}{k_\alpha E_e(k_\beta)} \right) N_{\beta \alpha},
\\
\check{N}_E^{\beta \beta}
&=&
\left( \frac{m_\beta E_i^\alpha}{k_\beta E_e^\alpha} \right)
\left( \check{I}_\alpha \check{I}_\beta + \check{I}_\alpha^2 \check{I}_\beta^2 + \check{I}_\alpha^3 \check{I}_\beta^3 + \cdots \right)
=
\left( \frac{m_\beta E_i^\alpha}{k_\beta E_e^\alpha} \right)
\left( \frac{\gamma_{\check{I}}^2}{1-\gamma_{\check{I}}^2} \right),
\label{eq_checkNbb}
%=
%\left( \frac{m_\beta E_i(k_\alpha)}{k_\beta E_e(k_\alpha)} \right) N_{\beta \beta}.
\end{eqnarray}
where
\begin{equation}
    \check{I}_\alpha = \check{k}_\alpha^{out} E_e^\beta,
    \quad
    \check{I}_\beta = \check{k}_\beta^{out} E_e^\alpha,
    \quad
    \gamma_{\check{I}} = \sqrt{\check{I}_\alpha \check{I}_\beta},
\end{equation}
and $\check{k}_\alpha^{out}$ and $\check{k}_\beta^{out}$ denote the average out-degrees of the species going to extinction.
% TS: I have learned to write in this way because "extinct" cannot be a verb. I would like to ask you for this grammatical thing.
%\kk{extincting species}.

\begin{figure}[tbhp]
\includegraphics[width=0.9\linewidth]{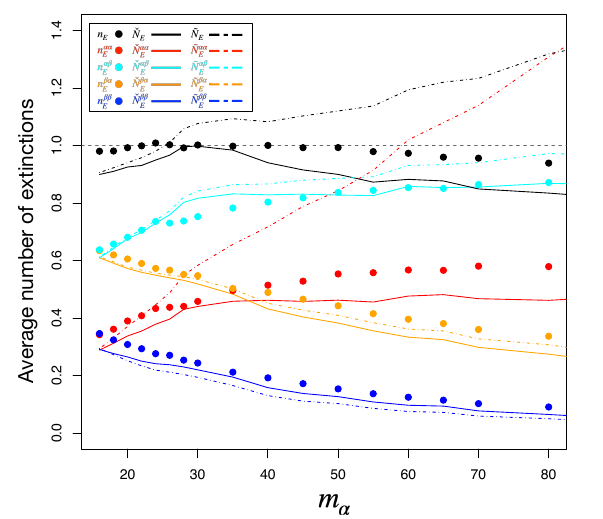}
\caption{
Expected numbers of extinctions $\check{N}_{\chi\eta}$ obtained using the extended %improved 
mean-field estimate with the observed average degrees $k_\chi$, the average out-degree of species going extinct in group $\beta$, and the extinction probability function $E(k)$ (Eqs.~\ref{eq_checkNaa}-\ref{eq_checkNbb}), with the initial degree $m_\beta$ fixed at $16$.
The improved estimate captures $\check{N}_{\chi\eta}$ for all $(\chi, \eta)$ pairs well, and hence the resulting average $\check{N}_E$ provides a quantitative explanation for the re-entrant transition to the growing phase at around $m_\alpha = 40$.
}
\label{fig_improved_mortalities}
\end{figure}

As shown by the solid lines in FIG.~\ref{fig_improved_mortalities}, the mean-field prediction with this correction successfully captures both the transition from the diverging phase to the finite phase and the subsequent re-entrant transition to diverging phase as $m_\alpha$ is increased from $16$ while keeping $m_\beta$
%with $m_\beta=16}
fixed at $16$.
%%%%% Degree of extinct species
%%

%\section{Discussion}
\section{Concluding Remarks}
For conventional unstructured systems, previous studies have established a phase transition from a diverging phase, in which the system size grows without bound, to a finite phase, in which it remains bounded, as the number of interactions increases. The present system with a bipartite graph structure exhibits an analogous transition.
However, when the initial degrees of newly introduced elements are asymmetric between the two partitions, the transition point shifts toward larger interaction numbers, indicating enhanced robustness against the addition of new elements.
This robustness becomes more pronounced with increasing asymmetry. Strikingly, the diverging phase persists even when the initial degrees on both sides exceed the transition point of the corresponding unstructured system. This strong effect due to
the asymmetric bipartite structure is naturally understood within a mean-field framework, previously successful for unstructured systems.
In the asymmetric case, the mean degree of emergent networks can shift significantly from the initial degree. In the part where the mean degree exceeds the initial degree, the increased mean degree suppresses the extinction probability while extinct elements remain strongly biased toward newly introduced ones and therefore retain a low average degree. The coexistence of these effects is the essential origin of the marked robustness in this regime.
These results reveal a simple and universal physical mechanism underlying robustness in evolving systems with asymmetric bipartite structure.

\bibliography{bipeos}
%%%
\end{document}